\documentclass[conference]{IEEEtran}

\usepackage[utf8]{inputenc}
\usepackage{times}
\usepackage{geometry}
\usepackage{graphicx}
\usepackage{cite}

\def\BibTeX{{\rm B\kern-.05em{\sc i\kern-.025em b}\kern-.08em
    T\kern-.1667em\lower.7ex\hbox{E}\kern-.125emX}}
\title{Vanlearning: A Machine Learning SaaS Application for People Without Programming Backgrounds}

\author{\IEEEauthorblockN{Chaochen Wu}
\IEEEauthorblockA{\textit{New York University} \\
cw2661@nyu.edu
}

}
\date{}
\begin{document}
\maketitle

\begin{abstract}
  Although we have tons of machine learning tools to analyze data, most of them require users have some programming backgrounds. Here we introduce a SaaS application which allows users analyze their data without any coding and even without any knowledge of machine learning. Users can upload, train, predict and download their data by simply clicks their mouses. Our system uses data pre-processor and validator to relieve the computational cost of our server. The simple architecture of Vanlearning helps developers can easily maintain and extend it.
\end{abstract}

\section{Introduction}
Currently, we employ machine learning algorithms to recognize people’s face, human’s voice, and means hidden in the text. However, some people who do not have any programming backgrounds missing their chance to learn the power of machine learning analysis. Therefore, we designed and implement a simple model to help people who have less programming analysis tabular data and visualize data online. Vanlearning is a simple software as a service (SaaS) system which aims to help people without any programming experience to understand the magic of machine learning algorithm to tabular data.

Several cloud machine learning services provide efficiently scalable high-performance resources for their customer \cite{pop2016machine}. Cloudnumbers.com allow users figure the size of clusters they want to use for their machine learning computation. CloudStat supports user analyze their data in R programming environment. Some cloud services providers, for example, Amazon \cite{AWS}, IBM \cite{IBM}, and Alibaba \cite{alibaba}, also provide machine learning services in the cloud. However, despite some of them provide comprehensible graph interface to reduce customers' programming workload, all of these cloud machine learning service providers requirement user have some programming skills, like Python or R, and have some primary knowledge for machine learning. Our system supports a "using first and understanding later" scenario help people who have absolutely zero backgrounds in programming and machine learning use and study basic machine learning algorithms. 

\section{Basic Architecture Design}
To design a simple machine learning application target to people without programming backgrounds we provide 3 general requirements systems functions. A SaaS application for machine learning need efficiently training user’s data and, for supervised learning, calculate output for test datasets. What’s more, for users without any programming backgrounds, it allows user analyze their data by a series of simple click and type actions. Finally, storing users’ information to ensure safety and convenience. 

\begin{figure}[htbp]
  \centerline{\includegraphics[width=\linewidth]{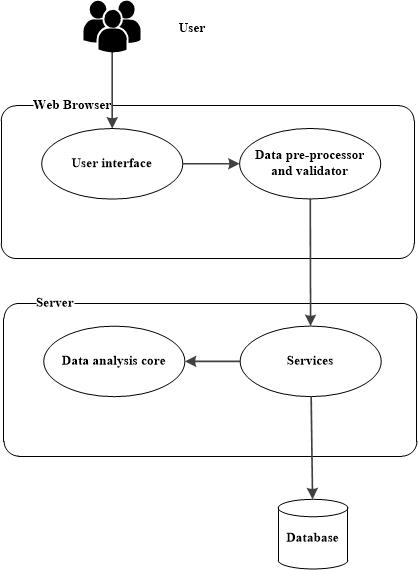}}
  \caption{Basic architecture of Vanlearning.}
  \label{fig1}
\end{figure}

Our application contains five parts. The user interface, which is the surface for user interacts with the software. Data pre-processor and validator, which process data and check data format and data quality. Services, which receive and response clients’ calling. The data analysis core which transforms original data to objects which can be analyzed and use machine learning libraries to analysis data. The database system, which stores all data for users' information and software operation. The basic architecture is showed in Fig.~\ref{fig1}.  
  
\subsection{User interface}
The user interface contains 3 parts, the navigation bar which allows user switch easily between different pages. The user information bar which contains user’s information and allows user sign out or any other operations. The canvas, which can be used for information display or as analysis dashboard. In addition, when a user visits the site at the first time without sign in, the page should prompt users to sign in or sign up. 

Canvas is the place where user read instruction or working on their data, a grid-like design can significantly decrease development workload and increase readability. In our system, we use boxes with different width and height to contain elements. For example, the box which takes up 100\% canvas can be used to display the large chunk of text, which includes software instruction and company’s information. Two boxes with 100\% width, one with 20\% height and one with 80\% height, with up and down arrangement can be used as dashboard and data table display. For an advanced system which allows users customize their canvas, for example, a user can add or delete operation box by mouse drag.

We use Gentelella \cite{gentelella} as a basic framework to implement our user interface. This framework provides mature support for login, navigation, and dashboard interfaces. Our system can rely on its windows which implemented by Bootstrap to hold our data table and operation buttons. To helping users inspect datasets they upload, we use a jQuery DataTables \cite{datatable} to implement tables contain users' datasets. Additional improvements can be allowing user directly modify their datasets in a table or allows user copy their datasets from their office software and paste them into our table.

\subsection{Data pre-processor and validator}
The data pre-processor and validator are a group of JavaScript programs which can process datasets from original CSV files, check the quality of data set, and parse data from JavaScript objects to JSON or XML for client-server communication. The data pre-processor and validator can significantly release the computational burden of server and prevent large chunks or illegal formats of data cause server crash. 

In the data pre-processor and validator, JavaScript programs will check whether the size or the format of use’s dataset is valid, if not it will forbid user send it to the server and prompt user upload new datasets. If a user wants to upload a file which contains non-numeric data and want to start an analysis, for example, linear regression, the JavaScript program with check the data of the CSV file and return an error after data check finish.

\subsection{Services}
The Services part of our application run on our server and response clients' messages. The business logic in services can be divided to two categories: user status, which contains user sign in, sign up, sign off, and sign in status check, this part interact frequency with the database. Data processing, which can parse user dataset, analysis data, check error and processing, and this part mainly interact with the data analysis core.

The Services build by Spring, Spring MVC, MyBatis, and Java 8. The DispatcherServlet, the central servlet of Spring MVC, provides a simple framework to receive and response client computer’s request \cite{gupta2010spring}.  The Controller module of Spring MVC handle HTTP GET or POST requests, and in Controller’s function, it packs data directly from clients to Plain old Java objects (POJO) which are called models in Spring MVC. For data analysis, The POJOs contains all required information, data, and parameters to finish a machine learning analysis, for example in k-means clustering analysis, the POJO of k-means contains cluster number, an integer class variable, and data, a String variable.

Beside process data from algorithm analysis, it also process data that need to interact with database: when user’s login data received from clients, it packages it to POJO and called login service to automatically create data access object (DAO) to search packed information from the database and return whether data from clients fit to record in database. The MyBatis framework map packed POJO to SQL for insert, search, or delete records \cite{mybatis}. Because MyBatis can use Java prepared statement to generate a template of SQL statements and replace it’s constant values when executed, it improves database safety by preventing SQL injection.

\subsection{Data analysis core}
Data analysis core in our system can be additionally separated to two parts, data parser and machine learning API. The format of input data is a String object which rows of original data separated by command, and individual row data represent by a JSON Object which its key is row name and value is data value in that row. We implement a parser to parse the string to an object which can directly feed to machine learning API with linear time complexity (O(n)). After we get the result from API, we use another parser to transform result to String and send back to clients' computers.

The machine learning API that we use is an open source machine learning library call Vanilla which is designed and implemented by ourselves.  It contains data structure to support 1 dimension and 2 dimension tensor operation. It implements several supervised learning and unsupervised learning algorithms, like logistic regression and k-means clustering. Despite we use data validators to prevent data in wrong formats is sent to our server, we employ try-catch blocks to catch exception generate by computation. Beside our own library, Weka \cite{hall2009weka}, a Java machine learning library, also can be used and support more machine learning algorithms.

\subsection{Database}
We use a simple relational database for storing information. The database in our system maintains user account information, which includes their username, password, and other personal information. We also provide user action storge to allow companies to analysis user actions and monitor user accounts. We also allow user can store their dataset in our database system, so users can open their datasets locate in our cloud. We implement our database by MySQL database. Because our database is relatively simple we only need 4 to 5 tables to support all business logic.

\section{Deployment}
\subsection{Deployment configuration}
The system currently deploys in Alibaba Cloud Elastic Compute service \cite{alibaba}. The hardware configuration is 1 core CPU, 1 GB memory, and 5Mbps bandwidth peak value. Our system is compiled and packed to a war file run in Java runtime environment 8, and in the cloud service, it deploys in Tomcat server 7.0.79 version and CentOS 7.3 64bit as the server and the operating system.

\subsection{Security}
As we mention before, for user information, we use MyBatis prepared statement to prevent SQL injections attack our database. Our cloud service provider also provides us with firewalls and services to prevent DDoS attacks. In addition, because machine learning algorithm takes up a lot of computation resources, to prevent robots attack our sign in/up and machine learning computation, we use Google recaptcha service \cite{recaptcha} to protect our system. If an IP is potentially be a robot the recaptcha requires it classify a group of image grids which are hard to robots before it can successfully finish an action.

\section{Evaluation}
\begin{table*}[htbp]
  \caption{Experiment results}
\begin{center}
  \begin{tabular}{ | c | c | c | c | c | c |}
  \hline
  \textbf{Module Name} & \textbf{Dataset} & \textbf{Training (sec)} & \textbf{Test (sec)} & \textbf{Parameters} & \textbf{Clicks}\\ 
  \hline
  k-means & Seeds & 1.1 & / & 1 & 1\\ 
  \hline
  Logistic regression & Haberman & 27.6 & 1.5 & 1 & 5\\ 
  \hline
  Linear Regression & Self-generated & 5.1 & 1.2 & 1 & 5\\ 
  \hline
  Decision Tree & Iris & 1.6 & 0.67 & 0 & 5\\ 
  \hline
  \end{tabular}
  \label{tab1}
\end{center}
\end{table*}
\subsection{Experiment setup}
To evaluate the performance of our system we use several public datasets from UCI machine learning repository \cite{Dua:2017} and self-generated data. In table 1 we show the dataset name, training time, test time, parameters number, and click times to finish a computation.

\subsection{Results}
Table~\ref{tab1} shows the experiment results. For the k-means clustering and the decision tree computation, our system can finish training within 2 seconds. The logistic regression and linear regression have longer computation time because we choose a small step size to get more accurate results with longer time \cite{barzilai1988two}. For prediction, all algorithms return the result within 1.5 seconds. In k-means clustering, users need to provide 1 parameters which are the number of clusters \cite{hartigan1979algorithm}. For logistic regression and linear regression, users need to tell our system which column of the table is input. The decision tree analysis does not need any parameters because it fixes the last column as output column because it allows user provides non-numeric data in the output column. For all supervised learning algorithms, users need to click 5 times of mouse to finish training and predict data; For k-means clustering, an unsupervised learning algorithm, users only need to click one time to get the result.

\section{An Example Scenario}
\begin{figure}[htbp]
  \centerline{\includegraphics[width=\linewidth]{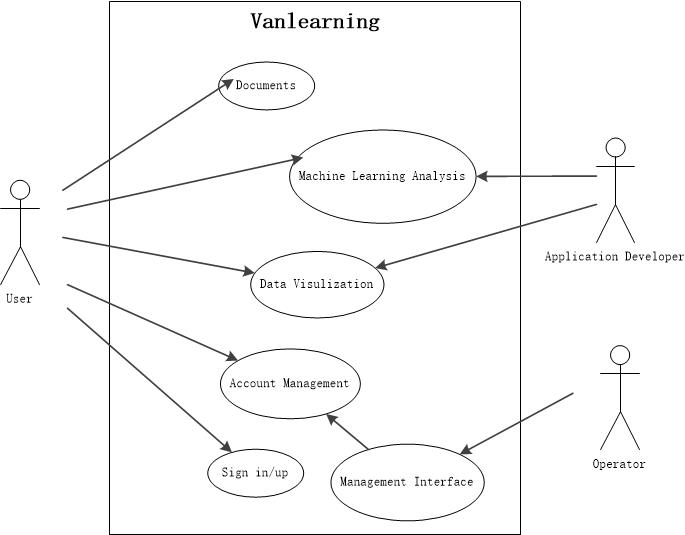}}
  \caption{Use Case Diagram for Vanlearning}
  \label{fig2}
\end{figure}

Fig.~\ref{fig2} shows a diagram for interactions between actors and our system. This scenario illustrates a new user who first time uses our system and finish his/her whole analysis work. At first, the user reads the instruction from our document and learning how to use our system and prepare valid data format. Then the user starts to use our system to analysis his/her data. The first step is to upload the dataset and start to training model, if choosing supervised learning, the user should also provide training dataset. After the user gets processed the dataset, he/she can download this dataset by various versions, like csv, pdf, and txt. If the user wants to visualize the processed data, he/she can also use our data visualization modules to look the dataset. The developer can easily add more modules on our basic system, like add more machine learning algorithms or add more data visualization tools. In addition, the system manager can get user action data from the database, and uses them to study user preference.

\section{Conclusion}
In this paper, we provide a new SaaS application which allows users without programming skills to analyze their datasets by machine learning algorithms. This system is simple and easy to expand, which help the company add more functions to meet different user needs. The structure of this system relieves the computational burden of machine learning, which allow company spend less on their computation resources. The simple and fluent user operation can attract more people to know the usages of machine learning.

To improve the performance of our system, we plan to optimize the machine learning core to provide a more efficient implementation of machine learning algorithms. What’s more, we will run more security tests to protect our system from web attacks. We will also provide more machine learning algorithms and add more data visualization tools to make our system solve more complicated problems.
\section*{Acknowledgment}
Thank Mr. Xin Nie and Mr. Zhitao Fan give me technical instructions. User icon made by Freepik from www.flaticon.com.
\bibliography{ref}
\bibliographystyle{ieeetr} 

\end{document}